\documentclass[aps,prl,twocolumn,10pt,nofootinbib]{revtex4-1}

\usepackage{amsmath,mathrsfs,graphicx}
\usepackage{multirow}

\begin{document}
\title{Cooperative Enhancement of Energy Transfer in a High-Density Thermal Vapor}
\author{L. Weller, R. J. Bettles, C. L. Vaillant, M. A. Zentile, R. M. Potvliege, C. S. Adams and I. G. Hughes} %
\affiliation{Joint Quantum Center (JQC) Durham-Newcastle, Department of Physics, Durham University, South Road, Durham, DH1 3LE, United Kingdom}
\date{\today}

\begin{abstract}
\noindent 
We  present an experimental study of energy transfer in a thermal vapor of atomic rubidium.  We measure the fluorescence spectrum in the visible and near infra-red as a function of atomic density using confocal microscopy.  At low density we observe energy transfer consistent with the well-known energy pooling process.  In contrast, above a critical density we observe a dramatic enhancement of the fluorescence from high-lying states that is not to be expected from kinetic theory.  We show that the density threshold for excitation on the D$_{1}$ and D$_{2}$ resonance line corresponds to the value at which the dipole-dipole interactions begins to dominate, thereby indicate the key role of these interactions in the enhanced emission.   
\end{abstract}
\maketitle
Thermal alkali atom vapors are finding an ever increasing range of applications including atomic clocks~\cite{knappe2005atomic}, magnetometry~\cite{budker2007optical}, electrometry~\cite{mohapatra2007coherent,bendkowsky2009observation}, quantum entanglement~\cite{julsgaard2004experimental} and memory~\cite{eisaman2005electromagnetically}, slow light~\cite{schmidt1996steep}, high-power lasers~\cite{rabinowitz1962continuous,krupke2003resonance}, frequency-up conversion~\cite{meijer2006blue}, narrowband filtering~\cite{abel2009faraday}, optical isolators~\cite{weller2012absolute,weller2012optical}, determination of  fundamental constants~\cite{truong2011quantitative} and microwave imaging~\cite{bohi2012simple}.  Some of these applications~\cite{vernier2010enhanced,meijer2006blue,olson2006two,akulshin2009coherent,shen2007many}, and the miniaturization of others, require relatively high atomic densities where dipole-dipole effects become important. Consequences of dipole-dipole interactions include self-broadening (see~\cite{weller2011absolute} and references therein), level shifts such as the Lorentz shift~\cite{maki1991linear,keaveney2012cooperative}, cooperative Lamb shift~\cite{keaveney2012cooperative}, determining the maximum refractive index in a gas~\cite{keaveney2012maximal} and intrinsic optical bistability~\cite{carr2013cooperative}. Although these phenomena have been studied extensively the system is sufficiently rich and complex that there are surprises still to be uncovered.  The dipole-dipole interaction is also important in other contexts such as non-radiative energy transfer~\cite{forster1948} and field enhancement in nanoplasmonics~\cite{vahid2002}.  One well-known phenomenon in alkali-metal vapors is energy pooling, which arises when two optically-excited atoms collide inelastically, resulting in energy transfer to states with higher energy.  Energy-pooling collisions have been studied extensively in sodium~\cite{kushawaha1980energy,bearman1978ionization}, potassium,~\cite{allegrini1982resonant,namiotka1997energy}, rubidium~\cite{barbier1999energy,yi2005energy,hill1982inelastic,caiyan2006studies,orlovsky2000theoretical,afanasiev2007laser,mahmoud2005electron,ban2004rb} and cesium~\cite{vadla1996energy,vadla1998energy,jabbour1996energy,gagne2002laser,de1999experimental,wang2002thermal,kang2002energy} and play an important role in frequency conversion schemes~\cite{meijer2006blue}.   

In this letter we report an  enhancement of energy transfer in Rb vapor greatly exceeding that expected from  energy pooling.  We study the evolution of the frequency up-converted fluorescence from highly-excited atoms driven on a resonance line with natural  linewidth $\Gamma$.  We show that energy transfer only occurs beyond a threshold density of the order of ${\cal N}_{\rm c}$ at which the dipole-dipole interaction becomes comparable to the natural broadening.  We define the critical density, ${\cal N}_{\rm c}$, as ${\cal N}_{\rm c}=\Gamma/\beta$, where $\beta$ is the dipole-dipole induced self-broadening parameter~\cite{lewis1980collisional,weller2011absolute}. For resonance lines with lower and upper state degeneracies $g_{\rm a}$ and $g_{\rm b}$ the  critical number density is ${\cal N}_{\rm c}=k^3/\left(2\pi\sqrt{g_{\rm b}/g_{\rm a}}\right)$, where $k$ is the wavevector of the excitation beam. For the Rb D$_{1}$ (5$^{2}$S$_{1/2}$ $\rightarrow$ 5$^{2}$P$_{1/2}$) and D$_{2}$ (5$^{2}$S$_{1/2}$ $\rightarrow$ 5$^{2}$P$_{3/2}$)  lines  these are  $7.9~\times~10^{13}~$cm$^{-3}$ and $5.9~\times~10^{13}~$cm$^{-3}$, easily obtained in thermal vapors.

\begin{figure}[b]
\centering
\includegraphics*[width=0.46\textwidth]{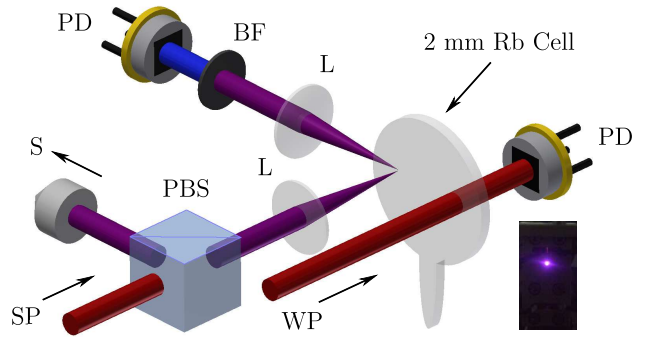}
\caption{Schematic of the confocal microscopy and side imaging experiments.  A linear strong-probe (SP) beam resonant with either the D$_{1}$ or D$_{2}$ lines transmitted by a polarizing beam splitter (PBS) is sent through a 2~mm thermal Rb vapor cell.  For confocal microscopy, a lens (L) is used to image the fluorescence onto a broadband multi-mode fiber connected to a spectrometer (S).  For side imaging, a blue bandpass filter (BF) is used to select the blue fluorescence from the cell, which is then incident on a photodiode (PD).  A weak-probe (WP) beam is used to measure the atomic density as a function of the oven temperature.  The inset shows a photograph of the fluorescence seen with a single input color.}
\label{Figure1}
\end{figure} 

\begin{figure*}[t]
\centering
\includegraphics*[width=1.0\textwidth]{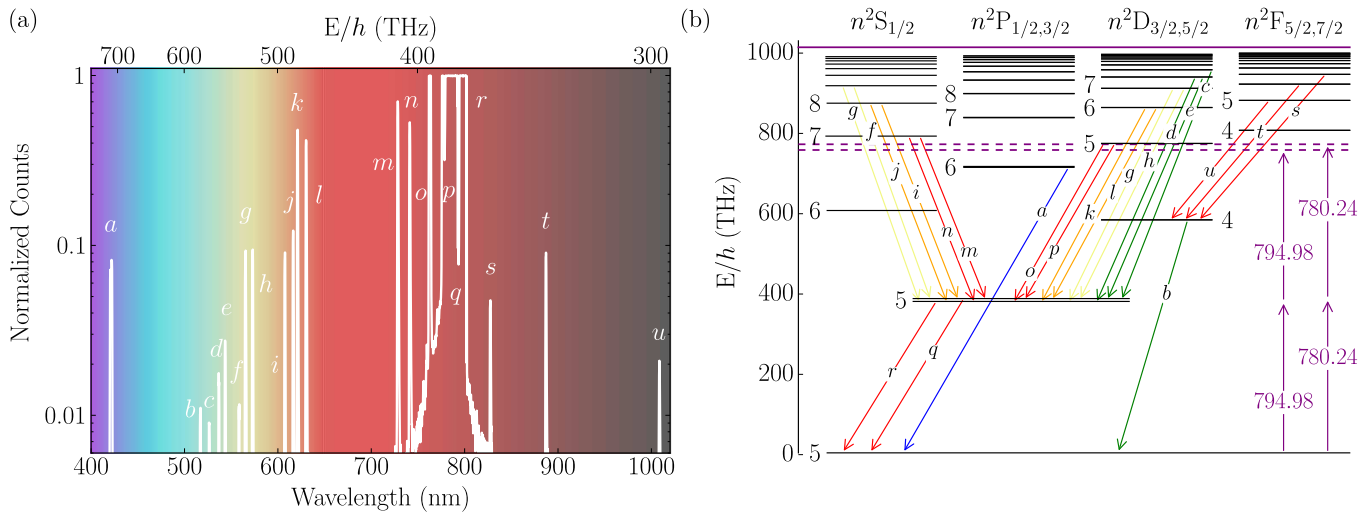}
\caption{(color online) Fluorescence spectrum and energy levels for Rb.  Panel~(a) shows the (solid white) normalized measured counts as a function of wavelength and frequency for input light resonant with the $F = 3 \rightarrow F'$ transition on the D$_{2}$ line of $^{85}$Rb.  This spectrum was measured for a number density of $9.0~\times~10^{14}~$cm$^{-3}$ (200$^{\circ}$C).  Panel~(b) shows the partial term diagram for a Rb atom.  The (dashed purple) quasi-two-photon resonances are for the 780.24~nm and 794.98~nm lines.  The (solid colored) lines labeled $a-u$ in (b) identify  the fluorescence lines measured in (a).}
\label{Figure2}
\end{figure*} 

The two experiments performed are  shown schematically in Fig.~\ref{Figure1}.  We sent light resonant with either the D$_{1}$ or D$_{2}$ lines through a 2~mm thermal vapor cell containing Rb atoms in their natural abundance (72$\%$ $^{85}$Rb, 28$\%$ $^{87}$Rb).  The Pyrex vapor cell was contained in an oven with the same design as in~\cite{weller2011absolute}. Beam widths of 1/e$^{2}$ radius 5.9~$\pm$~0.1~$\mu$m and 6.6~$\pm$~0.2~$\mu$m, with beam powers of 60~mW and 80~mW, gave maximum beam intensities of $1.1~\times~10^{5}$~Wcm$^{-2}$ and $1.2~\times~10^{5}$~Wcm$^{-2}$ for the D$_{1}$ and D$_{2}$ lines, respectively.  In one experiment we used confocal microscopy to image the fluorescence onto a broadband multi-mode fiber connected to a spectrometer, and in the other experiment we used side imaging to collect blue fluoresence.  The spectrometer has a bandwidth of $\sim$~700~nm over the visible and near infra-red, with a FWHM resolution of $\sim$~1.5~nm.

Fig.~\ref{Figure2} shows a typical fluorescence spectrum and the relevant energy levels.  Panel~(a) shows the normalized fluorescence counts as a function of wavelength and frequency for the resonant D$_{2}$ excitation of $^{85}$Rb.  The counts were normalized to the maximum (saturated) fluorescence of the 780.24~nm line.  We measured the spectrum for a number density of $9.0~\times~10^{14}~$cm$^{-3}$ (200$^{\circ}$C).  The  fluorescence lines correspond to the labeled transitions in panel~(b).  For an exposure time of 500~ms on the spectrometer all of the 21 lines are visible, however the large amount of resonant fluorescence bleaches near the excitation wavelength.  Similar spectra are measured for excitation resonant with the D$_{1}$ line.  Panel~(b) shows the partial term diagram for a single Rb atom showing the energy levels for the  orbital angular momentum states S, P, D and F. The quasi-two-photon resonances of 780.24~nm and 794.98~nm highlight the close proximity of the $5^{2}$D$_{3/2,5/2}$ energy levels.  The line at 1010.03~THz is the ionization limit.  The lines labeled $a-u$ show the allowed transitions\footnote{Note that $b$ ($4^{2}$D$_{3/2,5/2}$ $\rightarrow$ $5^{2}$S$_{1/2}$) is a quadrupole transition~\cite{hertel1969generalized} arising mainly due to decays from higher states into $4^{2}$D$_{3/2,5/2}$.}; further spectroscopic details  can  be found in table~\ref{Table1} of the supplemental material.  

It is evident that the observed spectrum requires population of excited states which cannot be accessed energetically from the sum of two resonant D-line photons.  From table~\ref{Table2}  and Fig.~\ref{Figure6} of the supplemental material we see that energy pooling can explain the origin of only some of the lines: the energy defect $\Delta$E between the initial and final pair states can be compensated kinematically.  By contrast the energy defect for excited states responsible for the majority of the lines are significantly larger than $k_\mathrm{B}T$; two colliding optically excited 5$^{2}$P$_{3/2}$ or 5$^{2}$P$_{1/2}$ atoms are unlikely to gain enough energy from translational motion to populate these high-lying states.  There must be another mechanism for populating these excited states; below we provide evidence that there is cooperative enhancement with dipole-dipole interactions playing a key role.

\begin{figure}[t]
\centering
\includegraphics*[width=0.46\textwidth]{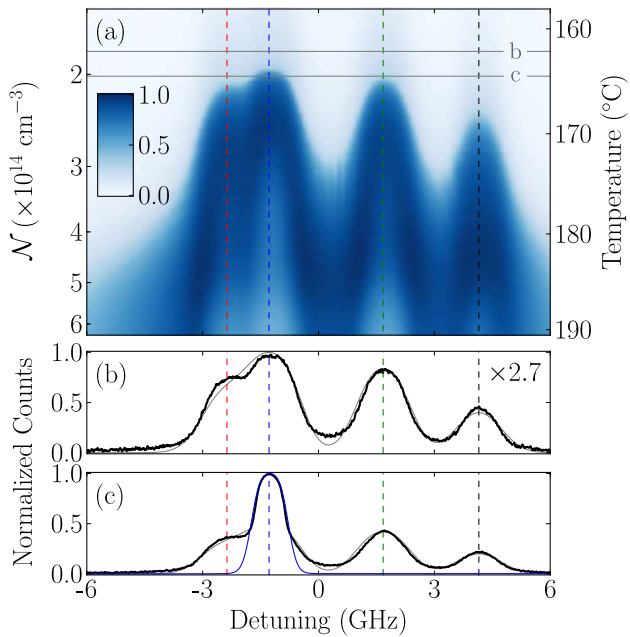}
\caption{(color online) (a) Fluorescence of the $6^{2}$P$_{1/2,3/2}$ $\rightarrow$ $5^{2}$S$_{1/2}$ transitions as a function of the number density (and corresponding temperature, see~\cite{weller2011absolute}) for input light scanned across the D$_{2}$ line.  The dashed (blue and green) lines correspond to the maximum fluorescence for the $F = 3 \rightarrow F'$ and $F = 2 \rightarrow F'$ transitions of $^{85}$Rb, respectively.  The dashed (red and black) lines correspond to the maximum fluorescence for the $F = 2 \rightarrow F'$ and $F = 1 \rightarrow F'$ transitions of $^{87}$Rb, respectively.  (b) and (c) show the measured normalized fluorescence for number densities and corresponding temperatures of $\mathcal{N} = 1.9~\times~10^{14}$~cm$^{-3}$ (163$^{\circ}$C) and $\mathcal{N} = 2.0~\times~10^{14}$~cm$^{-3}$ (165$^{\circ}$C), respectively.  A normalization factor $\times$~2.7 is used in (b).  The solid (grey) line in (b) and (c) has been calculated for a Doppler width of $2kv$.  The solid (blue) line in (c) has been calculated for Doppler widths of $kv$.}
\label{Figure3}
\end{figure}

We can eliminate the possibility of  mechanisms such as ionization or plasma formation by investigating the spectral dependence of the blue light (420~nm 6$^{2}$P$_{3/2}$ $\rightarrow$ 5$^{2}$S$_{1/2}$ and 422~nm 6$^{2}$P$_{1/2}$ $\rightarrow$ 5$^{2}$S$_{1/2}$, line $a$~Fig.\ref{Figure2}).  The cell heater allows optical access from the side (see Fig.~\ref{Figure1}) which we used to image the illuminated atoms along the length of the cell.  A lens  collimated the fluorescence onto a calibrated photodiode that has large gain over the visible regions.  A bandpass filter before the photodiode eliminated all other colors. The excitation laser was scanned over the Rb D$_2$ line,  and we used hyperfine/saturated absorption spectroscopy~\cite{smith2004role} to calibrate the detuning.  Fig.~\ref{Figure3} shows the spectral dependence of the measured fluorescence for the $6^{2}$P$_{1/2,3/2}$ $\rightarrow$ $5^{2}$S$_{1/2}$ transitions as a function of the number density.    Panels (b) and (c) are the spectra recorded for the number densities denoted by the solid lines b and c in panel (a).  The number densities  and corresponding temperatures are $\mathcal{N} = 1.9~\times~10^{14}$~cm$^{-3}$ (163$^{\circ}$C) and $\mathcal{N} = 2.0~\times~10^{14}$~cm$^{-3}$ (165$^{\circ}$C), respectively.  As the cell is heated it is evident that there is a very sudden and dramatic change in the fluorescence.
The production of blue light for the lowest density can be explained by the process of energy pooling: two excited ($5^{2}$P$_{3/2}$) atoms can produce a 5$^{2}$S$_{1/2}$ atom and a $5^{2}$D$_{3/2,5/2}$ atom with a positive energy defect. At 163$^{\circ}$C there is  enough thermal kinetic energy (at 163$^{\circ}$C $k_{\rm B}T$ = 9.1~THz) to compensate for the defect and populate $5^{2}$D$_{3/2,5/2}$ states; these then decay to 6$^{2}$P$_{1/2}$ from where the blue light is emitted. 

In order to understand the width of the spectral features we calculated  theoretical optical depth curves using the susceptibility model of~\cite{siddons2008absolute, weller2011absolute}, with a length scale of 15~$\mu$m.  This length scale was extracted from a fit to Fig.~\ref{Figure3}(b) and is fixed for the other data set.  The solid grey lines in Fig.~\ref{Figure3}(b) and (c) have a Doppler width of $2kv$; we see no evidence for the medium having become a hot plasma, nor  other possible population transfer mechanisms such as multi-photon ionization or dimer formation.  The solid blue line in Fig.~\ref{Figure3}(c) has a Doppler width of $kv$.  For this number density  for the $F = 3 \rightarrow F'$ in $^{85}$Rb transition  we observe enhanced blue fluorescence.   Taken together experiment and theory show a clear narrowing of the spectral dependence in the enhanced regime. These observations cannot be explained by energy pooling and require a different mechanism.  The four spectral lines lines are enhanced at different densities owing to the different isotopic abundances and state degeneracies for the  transitions.    Also note that for higher temperatures and densities radiation trapping is clearly visible on resonance.

\begin{figure}[t]
\centering
\includegraphics*[width=0.46\textwidth]{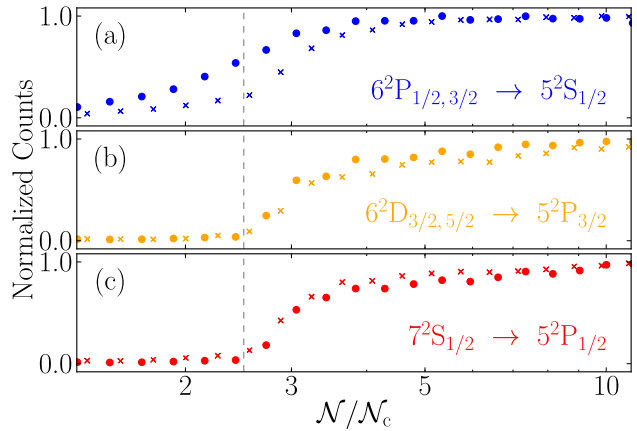}
\caption{(color online) Fluorescence as a function of the ratio of number density to the critical number density for the resonant D$_{1}$ (circles) and D$_{2}$ (crosses) excitation.  (a), (b) and (c)~show the normalized measured fluorescence for the 422~nm (blue), 630~nm (orange) and 728~nm (red) transitions, respectively.  For (a) the state can be accessed via the energy-pooling process with a positive energy defect.  We observe a monotonic increase in fluorescence as a function of density. For (b) and (c) the population of the excited state via energy pooling is kinematically suppressed as the energy defect is far larger than the collisional energy.  No signal is seen from these two lines until a dramatic turn on when the number density reaches  a threshold $\sim 2.5\,{\cal N}_{\rm c}$  for both D$_{2}$ and D$_{1}$ lines. The experimental uncertainties are approximately the size of the markers.}
\label{Figure4}
\end{figure}

A full analysis of the  blue-light generation is complicated by the fact that there is more that one mechanism which can populate the 6$^{2}$P states, and the medium is optically thick to blue light at the densities of interest.  So, to isolate the cooperative mechanism we  study the threshold behavior of spectral lines which do not originate from energy pooling.   These experiments illustrate the role of resonant dipole-dipole interactions in populating high-lying excited states. Fig.~\ref{Figure4} shows the normalized fluorescence as a function of the ratio of number density to the critical number density for resonant D$_{1}$ and D$_{2}$ excitation of $^{85}$Rb. We centered the frequencies on the $F = 3 \rightarrow F'$ transitions of $^{85}$Rb for the D$_{1}$ and D$_{2}$ lines.  For each measured value of the number density we extract an uncertainty from a least-squares fit using absolute absorption spectroscopy~\cite{weller2011absolute}.  Panel~(a) shows the normalized measured fluorescence at 422~nm   as a function of the ratio of number density to the critical number density.   Panels~(b) and (c) show the normalized measured fluorescence for the 630~nm (6$^{2}$D$_{3/2,\,5/2}$ $\rightarrow$ 5$^{2}$P$_{3/2}$) and 728~nm (7$^{2}$S$_{1/2}$ $\rightarrow$ 5$^{2}$P$_{1/2}$) transitions, respectively.  
Below the threshold density blue light can be produced via energy pooling.  By contrast no fluorescence is measured for the 630~nm  and 728~nm transitions within the sensitivity of our detector.  The energy defect for these high-$n$ transitions is too large for energy-pooling processes to be responsible for state transfer (table~\ref{Table2} of the supplemental material).  However fluorescence from these lines shows a dramatic increase for densities higher than a threshold number density for both D$_{1}$ and D$_{2}$ excitation. The threshold is  $\sim 2.5 \, {\cal N}_{\rm c}$  for both D$_{2}$ and D$_{1}$ lines. Recall that the ratio of ${\cal N}_{\rm c}$ for D$_{1}$ to D$_{2}$ lines is 1.33. We see similar  behavior for other lines arising from high-lying states with the enhancement always occurring at the same threshold density.  These results highlight the importance of the critical density (and thus dipole-dipole interactions) for  enhanced population transfer.  Note that at $\mathcal{N}/\mathcal{N}_{\rm c} = 1$ the atoms within the thermal vapor are separated by a distance  $\sim\lambda/2\pi$ for the D lines.  However, at this separation the resonant dipole-dipole interactions are  $\sim\Gamma/2\pi$, orders of magnitude smaller than the THz shifts needed to populate the high-lying states.  We now provide a plausible explanation for the dramatic enhancement of the population of high-lying states.

\begin{figure}[t]
\centering
\includegraphics*[width=0.46\textwidth]{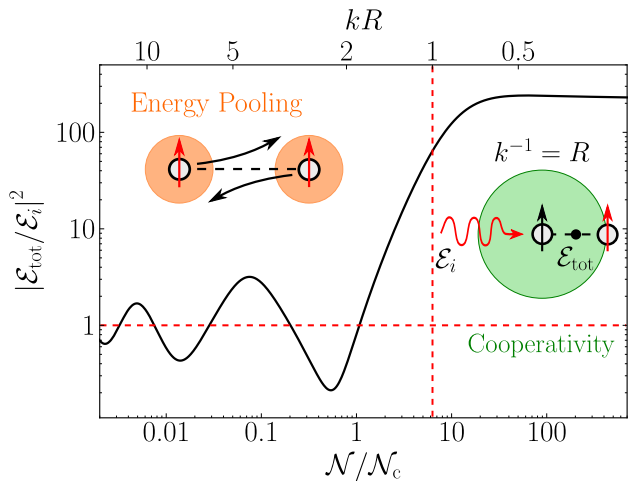}
\caption{(color online). Enhancement in the electric field midway between two atoms coupled by resonant dipole-dipole interactions as a function of the their separation.  The total electric field, $\mathcal{E}_{\rm tot}$, is calculated by summing the driving field, $\mathcal{E}_{\rm i}$, and the fields produced by each of the atomic dipoles.  The two atoms are analogous to the mirrors of an optical cavity, producing enhancement of the intensity of over two orders of magnitude for separations of $R < \lambda/2\pi$ (dotted vertical red line).   The number density axis is calculated assuming that $R = \mathcal{N}^{-1/3}$, and taking ${\cal N}_{\rm c}=k^3/2\pi$.}
\label{Figure5}
\end{figure}
 
We use a semi-classical model presented in~\cite{chomaz2012absorption} to calculate the total electric field midway between two atoms coupled by the resonant dipole-dipole interaction.   Fig.~\ref{Figure5} shows the intensity as a function of separation (correspondingly  number density).    For separations of $R < \lambda/2\pi$ we observe a dramatic enhancement in the total electric field analogous to field enhancement inside a resonant cavity. The threshold behavior of the  enhancement of the dipole-dipole interactions  suggests  a cooperative mechanism for the  energy transfer observed in our experiment.

In summary, we have extended the previous experimental studies of state transfer  in alkali-metal vapors to higher density and to include higher-lying states.  For blue-light production low-density transfer arises due to the well-known energy-pooling effect with collisions between two identical atoms, in their first excited states. Analysis of the blue fluorescence indicates a single-atom emission in the enhanced regime.  We have observed enhancement of energy transfer  for higher densities and attribute this to resonant dipole-dipole interactions in the cooperative regime. 
Above a threshold density we observed pronounced fluorescence from excited states which are not populated by energy pooling. This threshold density occurs when the dipole-dipole interaction becomes comparable to the natural linewidth for both D$_{1}$ and D$_{2}$ excitation. Beyond threshold there is a dramatic enhancement of the field trapped between two dipoles.
These observations open interesting prospects for exploiting dipole-dipole enhanced fields in non-linear and quantum optics, for example  in realizing a heralded two-photon source~\cite{willis2011photon}.

We thank L. Stothert and T. Bauerle for their contribution in the experimental measurements and J. Keaveney, S. Gardiner, A. Arnold and S. Franke-Arnold for stimulating discussions.  We acknowledge financial support from EPSRC (grants EP/H002839/1 and EP/F025459/1) and Durham University. The data presented in this paper are available upon request.    

\newpage

\appendix

\section{Supplemental Material}

\subsection{Energy levels}
Table~\ref{Table1} shows the measured and calculated fluorescence in a thermal Rb vapor.  Each transition or group of transitions is assigned a label $a-u$, which is associated with the lines in Fig.~2 of the main paper.  These energy levels were calculated from~\cite{sansonetti2006wavelengths}.

\begin{table*}[t]
\centering
\caption{Wavelengths and frequencies of the measured and calculated fluorescence as a consequence of the energy transfer in a thermal Rb vapor.  Each transition or group of transitions are assigned a label associated with the lines in Fig.~2 of the main paper.}
\begin{tabular}{crccc|crccc}
\hline
\hline
Label                & \multicolumn{2}{c}{Wavelength (nm)}        & $E/h$ (THz) & Transition                                       &
Label                & \multicolumn{2}{c}{Wavelength (nm)}        & $E/h$ (THz) & Transition                                       \\\hline
\multirow{2}{*}{$a$} & \multirow{2}{*}{$\quad \Bigr\{$} & 420.30  & 713.29      & 6$^{2}$P$_{3/2}$ $\rightarrow$ 5$^{2}$S$_{1/2}$  & 
\multirow{2}{*}{$l$} & \multirow{2}{*}{$\quad \Bigr\{$} & 630.00  & 475.85      & 6$^{2}$D$_{5/2}$ $\rightarrow$ 5$^{2}$P$_{3/2}$  \\   
                     &                                  & 421.67  & 710.96      & 6$^{2}$P$_{1/2}$ $\rightarrow$ 5$^{2}$S$_{1/2}$  &
                     &                                  & 630.09  & 475.79      & 6$^{2}$D$_{3/2}$ $\rightarrow$ 5$^{2}$P$_{3/2}$  \\
\multirow{2}{*}{$b$} & \multirow{2}{*}{$\quad \Bigr\{$} & 516.65  & 580.27      & 4$^{2}$D$_{3/2}$ $\rightarrow$ 5$^{2}$S$_{1/2}$  &
$m$                  &                                  & 728.20  & 411.69      & 7$^{2}$S$_{1/2}$ $\rightarrow$ 5$^{2}$P$_{1/2}$  \\ 
                     &                                  & 516.66  & 580.25      & 4$^{2}$D$_{5/2}$ $\rightarrow$ 5$^{2}$S$_{1/2}$  &
$n$                  &                                  & 741.02  & 404.57      & 7$^{2}$S$_{1/2}$ $\rightarrow$ 5$^{2}$P$_{3/2}$  \\                     
\multirow{2}{*}{$c$} & \multirow{2}{*}{$\quad \Bigr\{$} & 526.15  & 569.79      & 9$^{2}$D$_{5/2}$ $\rightarrow$ 5$^{2}$P$_{3/2}$  &
$o$                  &                                  & 762.10  & 393.38      & 5$^{2}$D$_{3/2}$ $\rightarrow$ 5$^{2}$P$_{1/2}$  \\
                     &                                  & 526.17  & 569.77      & 9$^{2}$D$_{3/2}$ $\rightarrow$ 5$^{2}$P$_{3/2}$  &
\multirow{2}{*}{$p$} & \multirow{2}{*}{$\quad \Bigr\{$} & 775.97  & 386.34      & 5$^{2}$D$_{5/2}$ $\rightarrow$ 5$^{2}$P$_{3/2}$  \\                        
$d$                  &                                  & 536.41  & 558.89      & 8$^{2}$D$_{3/2}$ $\rightarrow$ 5$^{2}$P$_{1/2}$  &
                     &                                  & 776.16  & 386.25      & 5$^{2}$D$_{3/2}$ $\rightarrow$ 5$^{2}$P$_{3/2}$  \\   
\multirow{2}{*}{$e$} & \multirow{2}{*}{$\quad \Bigr\{$} & 543.30  & 551.80      & 8$^{2}$D$_{5/2}$ $\rightarrow$ 5$^{2}$P$_{3/2}$  &
$q$                  &                                  & 780.24  & 384.23      & 5$^{2}$P$_{3/2}$ $\rightarrow$ 5$^{2}$S$_{1/2}$  \\                    
                     &                                  & 543.33  & 551.77      & 8$^{2}$D$_{3/2}$ $\rightarrow$ 5$^{2}$P$_{3/2}$  &
$r$                  &                                  & 794.98  & 377.11      & 5$^{2}$P$_{1/2}$ $\rightarrow$ 5$^{2}$S$_{1/2}$  \\          
$f$                  &                                  & 558.03  & 537.23      & 9$^{2}$S$_{1/2}$ $\rightarrow$ 5$^{2}$P$_{1/2}$  &
\multirow{3}{*}{$s$} & \multirow{3}{*}{$\quad \Biggr\{$}& 827.37  & 362.34      & 7$^{2}$F$_{5/2}$ $\rightarrow$ 4$^{2}$D$_{5/2}$  \\
\multirow{2}{*}{$g$} & \multirow{2}{*}{$\quad \Bigr\{$} & 564.93  & 530.67      & 7$^{2}$D$_{3/2}$ $\rightarrow$ 5$^{2}$P$_{1/2}$  &
                     &                                  & 827.37  & 362.34      & 7$^{2}$F$_{7/2}$ $\rightarrow$ 4$^{2}$D$_{5/2}$  \\                      
                     &                                  & 565.53  & 530.11      & 9$^{2}$S$_{1/2}$ $\rightarrow$ 5$^{2}$P$_{3/2}$  &
                     &                                  & 827.40  & 362.33      & 7$^{2}$F$_{5/2}$ $\rightarrow$ 4$^{2}$D$_{3/2}$  \\  
\multirow{2}{*}{$h$} & \multirow{2}{*}{$\quad \Bigr\{$} & 572.57  & 523.59      & 7$^{2}$D$_{5/2}$ $\rightarrow$ 5$^{2}$P$_{3/2}$  &
\multirow{3}{*}{$t$} & \multirow{3}{*}{$\quad \Biggr\{$}& 887.09  & 337.95      & 6$^{2}$F$_{5/2}$ $\rightarrow$ 4$^{2}$D$_{5/2}$  \\ 
                     &                                  & 572.62  & 523.55      & 7$^{2}$D$_{3/2}$ $\rightarrow$ 5$^{2}$P$_{3/2}$  &
					           &                                  & 887.09  & 337.95      & 6$^{2}$F$_{7/2}$ $\rightarrow$ 4$^{2}$D$_{5/2}$  \\      
$i$                  &                                  & 607.24  & 493.70      & 8$^{2}$S$_{1/2}$ $\rightarrow$ 5$^{2}$P$_{1/2}$  &
                     &                                  & 887.13  & 337.94      & 6$^{2}$F$_{5/2}$ $\rightarrow$ 4$^{2}$D$_{3/2}$  \\ 
$j$                  &                                  & 616.13  & 486.57      & 8$^{2}$S$_{1/2}$ $\rightarrow$ 5$^{2}$P$_{3/2}$  &
\multirow{3}{*}{$u$} & \multirow{3}{*}{$\quad \Biggr\{$}& 1007.80 & 297.47      & 5$^{2}$F$_{5/2}$ $\rightarrow$ 4$^{2}$D$_{5/2}$  \\                      
$k$                  &                                  & 620.80  & 482.91      & 6$^{2}$D$_{3/2}$ $\rightarrow$ 5$^{2}$P$_{1/2}$  &
					           &                                  & 1007.80 & 297.47      & 5$^{2}$F$_{7/2}$ $\rightarrow$ 4$^{2}$D$_{5/2}$  \\ 
					           &                                  &         &             &                                                  &
                     &                                  & 1007.85 & 297.46      & 5$^{2}$F$_{5/2}$ $\rightarrow$ 4$^{2}$D$_{3/2}$  \\                
\hline
\hline
\end{tabular}
\label{Table1}
\end{table*}
\subsection{Potential Energy Curves}
\begin{figure}[b]
\centering
\includegraphics*[width=0.48\textwidth]{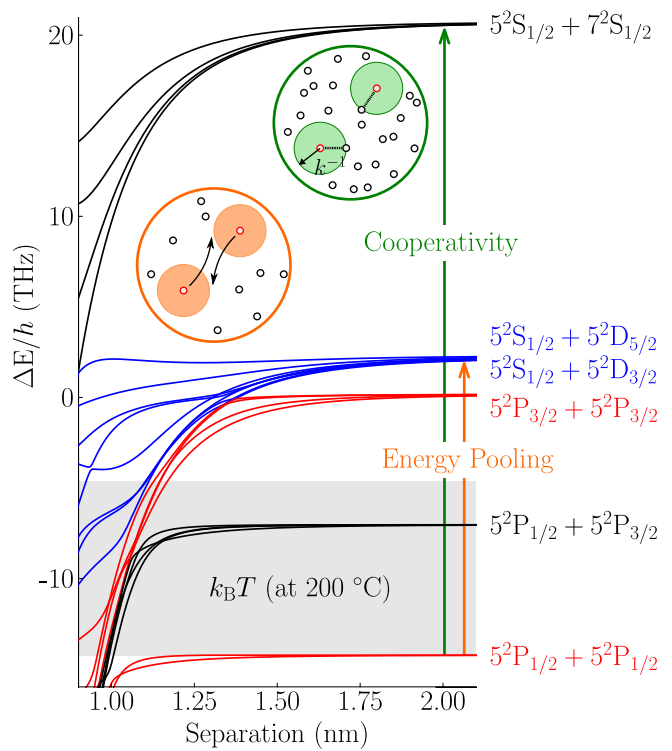}
\caption{(color online).  Potential energy curves for the $\Omega=0$ states of Rb around 5$^{2}$P$_{3/2}$ + 5$^{2}$P$_{3/2}$ resonance at close separation.  These resonances were calculated with a basis set of 500 states, including the ionic state curves.  The inset shows the two regimes of interest for state energy transfer.  In the low density regime energy pooling between two identical atoms in their first excited terms is the main process.  In the high density regime cooperative transfer for atoms in superpositions of the ground and excited terms is the dominant mechanism.}
\label{Figure6}
\end{figure}

The description of the energy pooling process depends on the calculation of dipole-dipole interactions.  This can be achieved using an effective Hamiltonian method, where the Born-Oppenheimer potential curves are given by diagonalizing the matrix, $H_{pq}= \delta_{pq}\epsilon_{p} + V_{pq}$.  The pair state energies, $\epsilon_{p}$, are given by the sum of the energies of the individual atomic states at infinite separation and the off-diagonal terms are given by
\begin{eqnarray}
V_{pq} = f(R)\frac{D_{11}}{R^3} \langle p_1 | r_1 | q_1 \rangle \langle p_2 | r_2 | q_2 \rangle~,
\label{effectivehamiltonian}
\end{eqnarray}
where $p$ and $q$ are state labels, $\langle p_i | r_i | q_i \rangle$ corresponds to the radial dipole matrix element for atom $i$, $R$ is the internuclear distance between the atoms and $D_{11}$ is a state-dependent angular factor given in~\cite{vaillant2012longrange}. The damping function, $f(R)$, accounts for the overlapping wavefunctions of the two electron clouds (ignoring exchange effects) given by~\cite{tang1984improved}
\begin{eqnarray}
f(R)= 1 - \exp^{-\gamma R}\sum_{l=0}^{3} \frac{(\gamma R)^l}{l!}~,
\label{dampingfunction}
\end{eqnarray}
where $\gamma$ is the repulsive range parameter.  The value of $\gamma$ is estimated for excited states by scaling the LeRoy radius, $R_{\mathrm{LR}} = 2(\langle r_1^2 \rangle ^{1/2} + \langle r_2^2 \rangle^{1/2})$, such that the ground state value matches the value given in~\cite{abrahamson1969born}. The radial dipole matrix elements are taken from precision calculations in the literature~\cite{safronova2011critically}, and supplemented using Coulomb wavefunctions~\cite{seaton2002coulomb} where literature values are unavailable.  Fig.~\ref{Figure6} shows the potential energy curves of Rb around the induced 5$^{2}$P$_{J}$ + 5$^{2}$P$_{J'}$ curves and resonant 5$^{2}$P$_{J'}$ + 5$^{2}$S$_{1/2}$ curves at close separation. Energy pooling arises from Landau-Zener type transitions between potential curves at avoided crossings.

\subsection{Kinetic Theory}
In an ideal gas at thermal equilibrium, the probability density for a single particle to have a (non-relativistic) speed, is given by the well known Maxwell-Boltzmann distribution~\cite{Blundell2009}.  However, for two particle collisions the distribution of relative speeds $v$ is a more useful quantity.  We can reduce this two-body problem to an effective one-body problem by simply replacing the mass $m$ of the individual particles with the reduced mass $\mu = m/2$.  From the Maxwell-Boltzmann distribution we find
\begin{eqnarray}
f_v(v) = 4\pi\left(\frac{m}{4\pi k_\mathrm{B}T}\right)^{3/2}v^2\exp\left(-\frac{mv^2}{4k_\mathrm{B}T}\right)~.
\label{eq:FinalApproachVelocity}
\end{eqnarray}
where $k_\mathrm{B}$ is the Boltzmann constant and $T$ is the absolute temperature.  When considering whether certain inelastic collisions are allowed by energy conservation, we want to compare the total energy before and after.  This is most convenient in the center of mass (CM) frame since the momentum sums to zero and so no energy is required to be carried away by kinetic energy after the process.

\begin{table*}[t]
\centering
\caption{Pair interactions for two excited Rb atoms ($5^{2}$P$_{1/2}$ + $5^{2}$P$_{1/2}$ + $\Delta$E$_{1}$ $\rightarrow$ States and $5^{2}$P$_{3/2}$ + $5^{2}$P$_{3/2}$ + $\Delta$E$_{2}$ $\rightarrow$ States.), highlighting the energy required in terms of THz and $k_\mathrm{B} T$ (at 200$^{\circ}$C = 9.86~THz) to reach the high-lying $n$ states.  The fraction of collisions, $\Delta\mathrm{F}$, is also shown.}  
\begin{tabular}{ccc|ccc|c}
\hline
\hline
\multicolumn{2}{c}{$\Delta$E$_{1}$} & \multirow{2}{*}{$\Delta\mathrm{F}$ (at 200$^{\circ}$C)} &	\multicolumn{2}{c}{$\Delta$E$_{2}$} & \multirow{2}{*}{$\Delta\mathrm{F}$ (at 200$^{\circ}$C)} & \multirow{2}{*}{States} \\ $\Delta$E$/h$ (THz) & $\Delta$E$/k_\mathrm{B} T$ (at 200$^{\circ}$C) &   & $\Delta$E$/h$ (THz) & $\Delta$E$/k_\mathrm{B} T$ (at 200$^{\circ}$C) & & \\\hline
-754.21							&	-76.50											&	1.00							  & -768.46             & -77.95									  & 1.00    	 					& $5^{2}$S$_{1/2}$ + $5^{2}$S$_{1/2}$ \\ 
-173.96							&	-17.65											&	1.00							  & -188.21             & -19.09									  & 1.00    	 					& $4^{2}$D$_{5/2}$ + $5^{2}$S$_{1/2}$ \\  
-173.94							&	-17.64											&	1.00							  & -188.19             & -19.09				         	  & 1.00       					& $4^{2}$D$_{3/2}$ + $5^{2}$S$_{1/2}$ \\
-150.65             & -15.28                      & 1.00      					& -164.90             & -16.73				            & 1.00       					& $6^{2}$S$_{1/2}$ + $5^{2}$S$_{1/2}$ \\
0.00                &	0.00												&	1.00								& -14.25			        & -1.44      								&	1.00			 					& $5^{2}$P$_{1/2}$ + $5^{2}$P$_{1/2}$ \\
7.12                &	0.72												&	0.70	    					& -7.12			          & -0.72											&	1.00			 					& $5^{2}$P$_{3/2}$ + $5^{2}$P$_{1/2}$ \\
14.25               &	1.44												&	0.41								& 0.00			          & 0.00											&	1.00			 					& $5^{2}$P$_{3/2}$ + $5^{2}$P$_{3/2}$ \\
16.27               &	1.65												&	0.35								& 2.02                & 0.20				              & 0.94       					& $5^{2}$D$_{3/2}$ + $5^{2}$S$_{1/2}$ \\
16.36               &	1.66												&	0.35								& 2.11                & 0.21					            & 0.93       					& $5^{2}$D$_{5/2}$ + $5^{2}$S$_{1/2}$ \\   
34.59               &	3.51												&	0.07								& 20.34               & 2.06					            & 0.25                & $7^{2}$S$_{1/2}$ + $5^{2}$S$_{1/2}$ \\
105.81              & 10.73                       & $8.43\times10^{-5}$ & 91.56               & 9.29					            & $3.35\times10^{-4}$ & $6^{2}$D$_{3/2}$ + $5^{2}$S$_{1/2}$ \\
105.88              & 10.74                       & $8.37\times10^{-5}$ & 91.63               & 9.29					            & $3.33\times10^{-4}$ & $6^{2}$D$_{5/2}$ + $5^{2}$S$_{1/2}$ \\   
116.52              & 11.82                       & $2.97\times10^{-5}$ & 102.27              & 10.37				              & $1.19\times10^{-4}$ & $8^{2}$S$_{1/2}$ + $5^{2}$S$_{1/2}$ \\  
153.56              & 15.58                       & $7.90\times10^{-7}$ & 139.31              & 14.13				              & $3.20\times10^{-6}$ & $7^{2}$D$_{3/2}$ + $5^{2}$S$_{1/2}$ \\  
153.61              & 15.58                       & $7.86\times10^{-7}$ & 139.36              & 14.14				              & $3.19\times10^{-6}$ & $7^{2}$D$_{5/2}$ + $5^{2}$S$_{1/2}$ \\ 
160.13              & 16.24                       & $4.14\times10^{-7}$ & 145.88              & 14.80				              & $1.68\times10^{-6}$ & $9^{2}$S$_{1/2}$ + $5^{2}$S$_{1/2}$ \\
181.78              & 18.44                       & $4.89\times10^{-8}$ & 167.53              & 16.99				              & $1.99\times10^{-7}$ & $8^{2}$D$_{3/2}$ + $5^{2}$S$_{1/2}$ \\
181.81              & 18.44                       & $4.87\times10^{-8}$ & 167.56              & 17.00			                & $1.99\times10^{-7}$ & $8^{2}$D$_{5/2}$ + $5^{2}$S$_{1/2}$ \\ 
199.78              & 20.26                       & $8.23\times10^{-9}$ & 185.53              & 18.82				              & $3.37\times10^{-8}$ & $9^{2}$D$_{3/2}$ + $5^{2}$S$_{1/2}$ \\
199.81              & 20.27                       & $8.21\times10^{-9}$ & 185.56              & 18.82				              & $3.36\times10^{-8}$ & $9^{2}$D$_{5/2}$ + $5^{2}$S$_{1/2}$ \\ 
\hline 
\hline
\end{tabular}
\label{Table2}
\end{table*}

Since the center of mass lies equidistant between the two particles of equal mass, we can immediately see that the speed of two colliding particles of equal mass should be $u_1'= u_2'= v/2$, where $u_1'$ and $u_2'$ are the CM frame speeds of particle 1 and 2, respectively.  From this we can derive an expression for the total kinetic energy, $E$, in the center of mass frame,
\begin{eqnarray}
\mathrm{E} = \frac{1}{2}m\,u_1'^2 +  \frac{1}{2}m\,u_2'^2 = m\left(\frac{v}{2}\right)^2~. 
\label{eq:KineticEnergy}
\end{eqnarray}
We now make use of the result from probability theory~\cite{Andrews2003}
\begin{eqnarray}
f_y(y)=f_x(x)\left|\frac{dx}{dy}\right|~,
\label{eq:PDF_oneRandomVariable}
\end{eqnarray}
where $y$ is some function of the random variable $x$ and $f_y$ and $f_x$ are the independent probability distributions of $y$ and $x$, respectively.  Eq.~(\ref{eq:PDF_oneRandomVariable}), along with eq.~(\ref{eq:FinalApproachVelocity}), allows us to arrive at the result
\begin{eqnarray}
f_\mathrm{E}(v) &=& \frac{8\pi}{m}\left(\frac{m}{4\pi k_\mathrm{B}T}\right)^{3/2}v\,\exp\left(-\frac{mv^2}{4k_\mathrm{B}T}\right)~, \nonumber \\
f_\mathrm{E}(\mathrm{E}) &=& 2\sqrt{\frac{\mathrm{E}}{\pi}}\left(\frac{1}{k_\mathrm{B}T}\right)^{3/2}\exp\left(-\frac{\mathrm{E}}{k_\mathrm{B}T}\right)~.
\label{eq:EnergyDistribution}
\end{eqnarray}

Integrating Eq.~\ref{eq:EnergyDistribution} with limits of $\mathrm{E}$ equal to the threshold energy $\Delta\mathrm{E}$ and infinity, will give the fraction of all collisions $\Delta \mathrm{F}$ that have at least $\Delta\mathrm{E}$ kinetic energy in the CM frame,
\begin{equation}
\Delta \mathrm{F} = \mathrm{erfc}\left(\sqrt{\frac{\Delta\mathrm{E}}{k_\mathrm{B}T}}\right) + 2\sqrt{\frac{\Delta\mathrm{E}}{\pi k_\mathrm{B}T}}\exp\left(\frac{-\Delta\mathrm{E}}{k_\mathrm{B}T}\right)~,
\label{eq:fractionOfCollisions}
\end{equation}
\begin{figure}[b!]
\centering
\includegraphics*[width=0.45\textwidth]{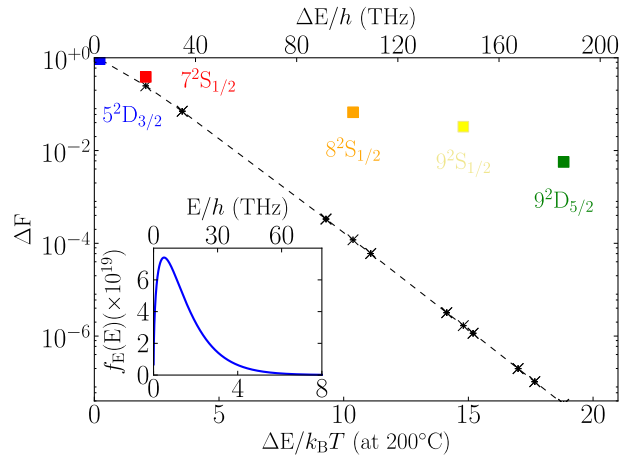}
\caption{(color online) Two particle kinetic energy distribution, $f_{\rm E}({\rm E})$, and fraction of collisions, $\Delta$F, with the threshold energy as a function of the energy in terms of THz and $k_\mathrm{B} T$ (at 200$^{\circ}$C). The squares are the measured normalied flourescence counts that we have attributed as being the result of population in the labeled states. For large energy defects $\Delta$E$/h$ = 185.56~THz ($5^{2}$P$_{3/2}$ + $5^{2}$P$_{3/2}$ $\rightarrow$ $9^{2}$D$_{5/2}$ + $5^{2}$S$_{1/2}$) one would expect eight orders of magnitude fewer collisions with enough energy for state transfer.  However only two orders of magnitude decrease in the amount of fluorescence is measured from the $9^{2}$D$_{5/2}$ state compared to the $5^{2}$D$_{3/2}$ state.}
\label{Figure7}
\end{figure}
where erfc denotes the complementary error function~\cite{Andrews2003}.  Fig.~\ref{Figure7} shows the two particle kinetic energy distribution and the fraction of collisions as a function of threshold energy in terms of THz and $k_\mathrm{B} T$ (at 200$^{\circ}$C).  The solid (black) crosses are the theoretical fraction of collisions calculated using Eq.~(\ref{eq:fractionOfCollisions}) and the energy defects $\Delta\mathrm{E}$ from table~\ref{Table2}.  The solid (coloured) squares are measured fluorescence taken from Fig.~2 of the main paper where the $5^{2}$D$_{3/2}$ state has been normalized to one.  In the inset the solid (blue) line shows the two particle energy distribution calculated using Eq.~(\ref{eq:EnergyDistribution}) at 200$^{\circ}$C.  In table~\ref{Table2} we list pair interactions for two excited Rb atoms, highlighting the energy required in terms of THz and $k_\mathrm{B} T$ (at 200$^{\circ}$C = 9.86~THz) to reach the high-lying $n$ states.  The fraction of collisions with enough energy, $\Delta\mathrm{F}$, are also shown.  At any given moment in a gas, the constituent particles will have varying distances to their nearest neighbor.  This is the case even if the overall number density, $\mathcal{N}$ is constant.  Assuming particles are placed randomly and are non-interacting, we get the following probability distribution for the distance, $R$, to the nearest neighbor~\cite{Chandrasekhar1943},
\begin{eqnarray}
P(R)=4\pi\mathcal{N}R^2\exp\left(-\frac{4\pi \mathcal{N}}{3}R^3\right)~.
\label{eq:nearestNeighbor}
\end{eqnarray}
The mean of this distribution is approximately at $\frac{5}{9}\mathcal{N}^{-1/3}$~\cite{Chandrasekhar1943}.   

\bibliographystyle{apsrev4-1}
\bibliography{myreferences}
\end{document}